\documentstyle[twocolumn,aps,epsf]{revtex}
\def\r{{\bf{r}}}
\def\a{{\bf{a}}}
\def\g{{\bf{g}}}

\def\k{{\bf{k}}}
\def\kt{{\tilde{\k}}}
\def\K{{\bf{K}}}
\def\M{{\bf M}}

\def\vereq#1#2{\lower3pt\vbox{\baselineskip1.5pt \lineskip1.5pt
\ialign{$#1\hfill##\hfil$\crcr#2\crcr\sim\crcr}}}

\begin{document}

\title{Systematic and Causal Corrections to the Coherent Potential 
Approximation}
\author{M. ~Jarrell$^1$, and H. R. Krishnamurthy $^2$}
\address{$^1$ Department of Physics, University of Cincinnati, Cincinnati,
OH 45221}
\address{$^2$ Department of Physics, IISc, Bangalore 560 012, India} 
\date{\today}
\maketitle

\begin{abstract}

	The Dynamical Cluster Approximation (DCA) is modified to include
disorder.    The DCA incorporates non-local corrections to local 
approximations such as the Coherent Potential Approximation (CPA) by 
mapping the lattice problem with disorder, and in the thermodynamic limit, 
to a self-consistently embedded finite-sized cluster problem.  It satisfies 
all of the characteristics of a successful cluster approximation.  It is 
causal, preserves the point-group and translational symmetry of the 
original lattice, recovers the CPA when the cluster size equals one, and 
becomes exact as $N_c\to\infty$.  We use the DCA to study the Anderson 
model with binary diagonal disorder.  It restores sharp features and band 
tailing in the density of states which reflect correlations in the local 
environment of each  site.  While the DCA does not describe the localization
transition,  it does describe precursor effects of localization.

\end{abstract}

\section{Introduction}

The Coherent Potential Approximation (CPA)\cite {cpa12,ekl74,agonis}
is a widely used method for treating disordered systems. Within the 
CPA, the problem is first averaged over all possible disorder 
configurations in an attempt to regain the translational invariance 
lost due to disorder. Then non-local correlations of the disorder 
potential are neglected leading to a self consistent single-site 
approximation  (like the Weiss mean field theory of magnetism).  
It has been applied with great success to a variety of 
problems\cite {cpa12,ekl74,agonis}, including {\em{ab-initio}} 
calculations in disordered metallic alloys\cite{kkrcpa}.

Nevertheless, the CPA fails to provide a completely satisfactory theory 
for disordered systems\cite{agonis}.  As a single-site mean field theory, 
it cannot account for the disorder induced, short ranged but nonlocal, 
correlations due to the local order in the environment of each site
responsible for band tailing and sharp structures in the density of
states. There have been many attempts\cite{mlsrv78,kldg80,agonis} 
to formulate non-local corrections to the CPA in which the lattice is 
mapped onto a self-consistently embedded finite-sized cluster.  However, 
as argued clearly and in detail by Gonis \cite{agonis}, these theories 
all fail in some significant way.  A successful theory must be able to 
account for fluctuations in the local environment in a self-consistent way,
become exact in the limit of large cluster sizes, and recover the CPA
when the cluster size equals one.  It must be easily implementable
numerically and preserve the translational and point-group symmetries 
of the lattice.  Finally, and most significantly, it should be fully causal 
so that the single-particle Green function and self energy are analytic 
in the upper half plane.  No presently existing cluster extension of the 
CPA satisfies {\em{all these requirements}}\cite{agonis,fntca}.

	Recently, a new method called the The Dynamical Cluster
Approximation  (DCA)\cite{DCA_hettler1,DCA_hettler2,DCA_maier1} was
developed for ordered correlated systems such as the Hubbard model to add 
non-local corrections to the Dynamical Mean Field Approximation. In this
manuscript, we modify the DCA to include disorder and show that the 
resulting formalism satisfies all of the above stated requirements for 
a successful cluster extension of the CPA.

	In the next Section, we review the basic diagrammatic perturbation 
theory formalism for disordered systems.  In Sec.~\ref{CPA} we show 
that the CPA is equivalent to neglecting momentum conservation at all 
internal vertices of the diagrams, and in Sec.~\ref{DCA} we introduce 
the DCA for disordered systems which systematically  restores the momentum 
conservation relinquished by the CPA.  In  Sec.~\ref{characteristics} we 
show that the DCA satisfies each of the desired  characteristics described 
above.  In Sec.~\ref{results} we show  results for the two-dimensional 
Anderson model with binary diagonal disorder.  Finally in Appendix A we 
present an alternate way of viewing and justifying the disordered DCA 
algorithm developed in this paper, using the replica (or other) methods 
of disorder averaging.

\section{Basic Formalism}  
\label{formalism}

	We consider an Anderson model with diagonal disorder, 
described by the Hamiltonian
\begin{equation}
H =
\displaystyle - \sum_{<ij>,\sigma} 
t \left( 
C^{\dagger}_{i,\sigma} C_{j,\sigma} + 
C^{\dagger}_{j,\sigma} C_{i,\sigma}
\right)  \\[5mm]
\displaystyle+
 \sum_{i \sigma}(V_i- \mu) n_{i,\sigma}
\end{equation}
where $C^{\dagger}_{i,\sigma}$ creates a quasiparticle on site $i$ with 
spin $\sigma$, $n_{i,\sigma}=C^{\dagger}_{i,\sigma} C_{i,\sigma}$.  The 
disorder occurs in the local orbital energies
$V_i$, which we assume are independent quenched random variables
distributed  according to some specified probability distribution  $P(V)$. 
{\em{The DCA formalism that we develop in this paper is a general method 
valid for any $P(V)$. }} However, for illustrative purposes, for the
specific calculations presented in this paper we take  $V_i=\pm V$ with
equal probability $1/2$ (binary disorder).   

\begin{figure}
\epsfxsize=3.3in
\epsffile{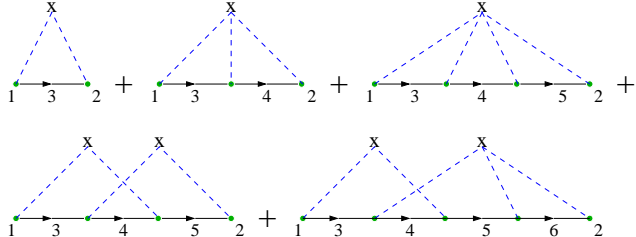}
\caption[a]{\em{The first few graphs in the irreducible self energy of a 
diagonally disordered system.  Each $\circ$ represents the scattering of 
a state $\k$ from sites (marked $X$) with a 
local disorder potential  
distributed  according to some specified probability distribution  $P(V)$.  
The numbers label the $\k$ states.}}
\label{dis_dia}
\end{figure}

The effect of the disorder potential $\sum_{i \sigma}V_i n_{i,\sigma}$ 
can be described using standard diagrammatic perturbation theory (although
we will  eventually sum to {\em{all}} orders).  It may be re-written in
reciprocal space as
\begin{equation}
H_{dis} = \frac{1}{N} \sum_{i,\k,\k',\sigma} V_i
C^{\dagger}_{\k,\sigma} C_{\k',\sigma} e^{i\r_i(\k-\k')}
\end{equation}
The corresponding irreducible (skeletal) contributions to the self energy 
may be represented diagrammatically\cite{agonis}
and are displayed in Fig.~\ref{dis_dia}.
\vspace{0.15in}
Here each $\circ$ represents the scattering of an electronic Bloch state
from a local disorder potential 
at some site represented by $X$. 
The  dashed lines connect  
scattering events that involve
the same local potential.    In each graph, the sums over the sites are
restricted so that 
the different $X$ 's  represent 
scattering from {\em{different}}
sites. No graphs representing  a single scattering event are included
since these may simply be absorbed  as a renormalization of 
the chemical potential $\mu$.

	Translational invariance and momentum conservation are restored 
by averaging over all possible values of the disorder potentials $V_i$.  
For example, consider the second diagram in Fig.~\ref{dis_dia}, given by 
\begin{equation}
\frac{1}{N^3}\sum_{i,\k_3, \k_4} \langle V_i^3 \rangle
G(\k_3)G(\k_4) e^{i\r_i\cdot(\k_1-\k_3+\k_3-\k_4+\k_4-\k_2)}
\end{equation}
where $G(\k)$ is the disorder-averaged single-particle Green function for 
state $\k$.  The average over the distribution of scattering potentials 
$\langle V_i^3 \rangle = \langle V^3 \rangle$ independent of i. 
After summation over the remaining labels, this becomes
\begin{equation}
\langle V^3 \rangle G(\r=0)^2 \delta_{\k_1,\k_2}
\end{equation}
where $G(\r=0)$ is the local Green function.  Thus the second diagram's 
contribution to the self energy involves only local correlations. Since 
the internal momentum labels always cancel in the exponential, the same 
is true for all non-crossing diagrams shown in the top half of 
Fig.~\ref{dis_dia}.
  
Only the diagrams with crossing dashed lines are non-local.  Consider the 
fourth-order diagrams such as those shown on the bottom left and upper 
right of Fig.~\ref{dis_dia}.  When we impurity average, we generate 
potential terms $\langle V^4 \rangle$ when the scattering occurs from the 
same local potential (i.e.\ the third diagram) or $\langle V^2 \rangle^2$ 
when the scattering occurs from different sites, as in the fourth diagram.  
When the latter diagram is evaluated, to avoid overcounting, we need to 
subtract a term proportional to $\langle V^2 \rangle^2$ but corresponding 
to scattering from the same site.  This term is needed to account for the 
fact that the fourth diagram should really only be evaluated for sites 
$i \neq j$! For example, the fourth diagram yields
\begin{eqnarray}
\large\langle \frac{1}{N^4} \sum_{i\neq j \k_3 \k_4 \k_5} 
V_i^2 V_j^2
e^{i\r_i\cdot(\k_1+\k_4-\k_5-\k_3)}
e^{i\r_j\cdot(\k_5+\k_3-\k_4-\k_2)}\nonumber\\
G(\k_5)G(\k_4)G(\k_3)
\large\rangle
\end{eqnarray}
Evaluating the disorder average $\langle\rangle$, 
we get the following two terms:
\begin{eqnarray}
\frac{1}{N^4} \sum_{i j \k_3 \k_4 \k_5} 
\langle V^2 \rangle^2
e^{i\r_i\cdot(\k_1+\k_4-\k_5-\k_3)}e^{i\r_j\cdot(\k_5+\k_3-\k_4-\k_2)}\nonumber\\
G(\k_5)G(\k_4)G(\k_3) \nonumber\\
-
\frac{1}{N^4} \sum_{i \k_3 \k_4 \k_5} 
\langle V^2 \rangle^2 e^{i\r_i\cdot(\k_1-\k_2)}
G(\k_5)G(\k_4)G(\k_3)
\end{eqnarray}
Momentum conservation is restored by the sum over $i$ and $j$; i.e.~over all 
possible locations of the two scatterers.  It is reflected by the Laue 
functions, $\Delta=N\delta_{\k+\cdots}$, within the sums
\begin{eqnarray}
\frac{\delta_{\k_2,\k_1}}{N^3} \sum_{\k_3 \k_4 \k_5} 
\langle V^2 \rangle^2
N\delta_{\k_2+\k_4,\k_5+\k_3}\nonumber\\
G(\k_5)G(\k_4)G(\k_3)\nonumber \\
-
\frac{\delta_{\k_2,\k_1}}{N^3} \sum_{\k_3 \k_4 \k_5} 
\langle V^2 \rangle^2
G(\k_5)G(\k_4)G(\k_3) 
\label{eval_dia4}
\end{eqnarray}
Since 
the first term in Eq.~\ref{eval_dia4} involves convolutions of $G(\k)$ it 
reflects non-local  correlations. [Local contributions such as the second 
term in Eq.~\ref{eval_dia4} can, if one so chooses, be combined together with 
the contributions from the corresponding local diagrams such as the third
diagram in Fig.~\ref{dis_dia}  by replacing $\langle V^4 \rangle$ in the
latter by the cumulant $\langle V^4 \rangle - \langle V^2 \rangle^2$ .]
Given the fact that different $X$'s must correspond to different sites, it 
is easy to see that all crossing diagrams must involve non-local correlations.

\section{Cluster Approximations}

\subsection{The Coherent Potential Approximation} 
\label{CPA}

\begin{figure}
\epsfxsize=3.3in
\epsffile{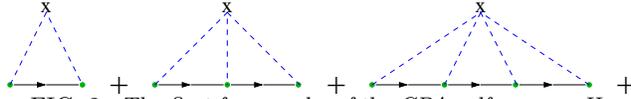}
\caption[a]{\em{The first few graphs of the CPA self energy.  Here the
Green function line represents the average local propagator.}}
\label{cpa_dia}
\end{figure}

In the Coherent Potential Approximation (CPA), nonlocal correlations 
involving different scatterers are ignored.  Thus, in the calculation of 
the self energy, we ignore all of the crossing diagrams shown on the bottom 
of Fig.~\ref{dis_dia}; and retain only the class of diagrams such as those 
shown on the top representing scattering from a single local disorder 
potential.  These diagrams are shown in Fig.~\ref{cpa_dia}.  
\vspace{0.15in}

Employing diagrammatic arguments, it is easy to see that the CPA is fully 
equivalent to the neglect of momentum conservation at each internal 
vertex.  This is accomplished by setting each Laue function within the 
sum (eg., in Eq.~\ref{eval_dia4}) to 1.  We may then freely sum over the 
internal momenta, leaving only local propagators.  All non-local self 
energy contributions (crossing diagrams) must then vanish.  For example, 
consider again the fourth graph. If we replace the Laue function 
$N\delta_{\k_1+\k_4,\k_5+\k_3}\to 1$ in Eq.~\ref{eval_dia4}, then the 
two contributions cancel and this diagram vanishes.  Thus an alternate 
definition of the CPA\cite{MH}, in terms of the Laue functions $\Delta$, 
is
\begin{equation}
\Delta = \Delta_{CPA} = 1
\end{equation}
I.e., the CPA is equivalent to the neglect of momentum conservation
at all internal vertices of the disorder-averaged irreducible graphs.

\subsection{The Dynamical Cluster Approximation}
\label{DCA}

\begin{figure}[ht]
\epsfxsize=2.7in
\epsffile{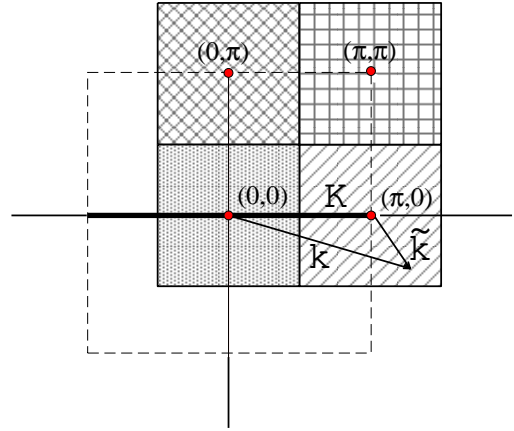}
\caption{{$N_c=4$ cluster cells (shown by different fill patterns) that 
partition the first Brillouin Zone (dashed line).  Each cell is centered
on a cluster momentum $\K$ (filled circles). To construct the DCA cluster, we 
map a generic momentum in the zone such as $\k$ to the nearest cluster point 
$\K=\M(\k)$ so that $\kt=\k-\K$ remains in the cell around $\K$.
} }
\label{brillouin}       
\end{figure}
	The DCA systematically restores the momentum conservation at internal
vertices which was relinquished by the CPA, and so incorporates non-local 
corrections.  This is done by dividing the Brillouin zone into $N_c$ equal 
cells, as shown in Fig.~\ref{brillouin} and requiring that momentum be 
partially conserved for momentum transfers between the coarse graining cells 
shown in Fig.~\ref{brillouin}, but ignored for momentum transfers within each 
cell.  This may be accomplished by employing the Laue functions
\begin{equation}
\Delta = \Delta_{DCA} = 
N_c \delta_{\M(\k_1)+\M(\k_2),\M(\k_3)+\M(\k_4)  ...} \label{Laue_DCA}
\end{equation}
where $\M(\k)$, illustrated in Fig~\ref{brillouin}, maps the momenta $\k$
to the nearest cluster momenta $\K$.  $\Delta_{DCA}$ becomes one when $N_c=1$
since then all momenta are mapped to the zone center by $\M$. Thus the
CPA is recovered in this limit.  Furthermore, as $N_c$ becomes large,
the exact result is recovered since $\lim_{N_c\to\infty} \M(\k)=\k$ for
all momenta $\k$. 
\begin{figure}[htb]
\epsfxsize=3.4in
\epsffile{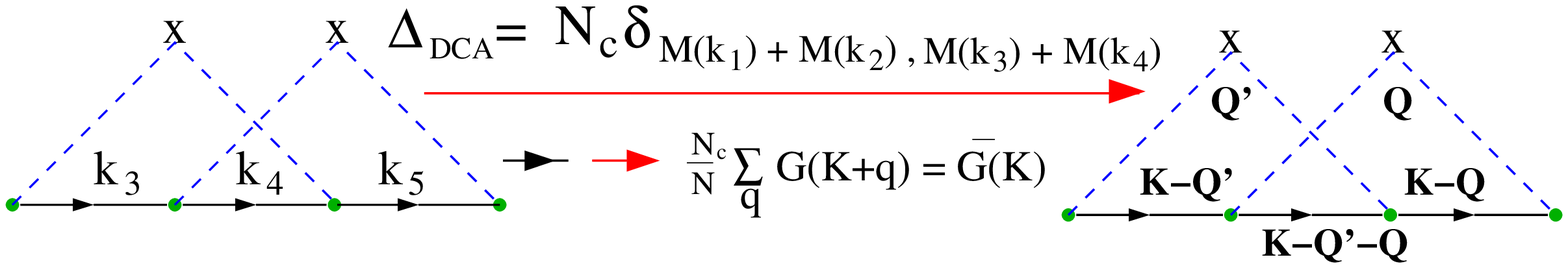}

\caption{Use of the DCA Laue function $\Delta_{DCA}$ leads to a 
leads to the replacement of the lattice propagators $G({\bf k}_1)$, 
$G({\bf k}_2)$, ... by coarse grained propagators $\bar{G}({\bf K})$, 
$\bar{G}({\bf K}^\prime)$, ...  }
\label{collapse_dis}
\end{figure}
If we employ the DCA Laue function in each of the self energy diagrams 
shown in Fig.~\ref{dis_dia} then we may freely sum over the momenta 
within each coarse-graining cell shown in Fig.~\ref{brillouin}.  As 
illustrated in Fig.~\ref{collapse_dis} for a fourth-order graph, this 
leads to the replacement of the lattice propagators $G({\bf k}_1)$, 
$G({\bf k}_2)$, ... by coarse grained propagators $\bar{G}({\bf K})$, 
$\bar{G}({\bf K}^\prime)$, .... which are given by 
\begin{equation}
\bar{G}(\K) \equiv \frac{N_c}{N}\sum_{\kt}G(\K+\kt),
\label{gbar}
\end{equation}
where $N$ is the number of points of the lattice, $N_c$ is the number of 
cells in the cluster, and the $\kt$ summation runs over the momenta of 
the cell about the cluster momentum  $\K$ (cf. Fig.~\ref{brillouin}).
	
As $N_c$ increases from one, systematic non-local corrections to the CPA 
self energy are introduced.  To see this, recall that the self energy 
is a functional of the Green function.  According to Nyquist's sampling 
theorem\cite{nyquist}, to reproduce correlations of length $\alt L/2$ in 
the Green function and corresponding self energy, we only need to sample 
the reciprocal space at intervals of $\Delta k\approx 2\pi/L$.  Knowledge 
of these Green functions on a finer scale in momentum is than $\Delta k$ 
unnecessary, and may be discarded to reduce the complexity of the problem.   
Thus the cluster self energy will be constructed from the {\em{coarse-grained 
average}} of the single-particle Green function within the cell centered 
on the cluster momenta. For short distances $r\alt L/2$, where $L$ is now
the linear size of the cluster, the Fourier transform of the Green function 
$\bar{G}(r) \approx G(r) +{\cal{O}}((r\Delta k)^2)$, so that short ranged 
correlations are reflected in the irreducible quantities constructed from 
$\bar{G}$; whereas, longer ranged correlations $r>L/2$ are cut off by the 
finite size of the cluster \cite{DCA_hettler2}.  

	We show in the appendix that that free-energy arguments presented
previously\cite{DCA_hettler2} apply to the disordered case as well.  In 
particular, the DCA estimate of the lattice free energy is minimized by 
the choice $\Sigma(\k,\omega)={\bar{\Sigma}}(\M(\k),\omega)$.

\paragraph*{Algorithm}
With the substitution $\Delta \to \Delta_{DCA}$, most of the diagrams 
represented in Fig.~\ref{dis_dia} remain. 
However, the complexity of the problem is greatly reduced since 
the nontrivial sums involve only the cluster momenta $K$ ( numbering $N_c$
instead of $N$).  Furthermore, since these  diagrams are the same as those
from a finite-sized periodic cluster of $N_c$  sites, we can easily sum
this series to all orders by numerically solving  the corresponding
cluster problem.  The resulting algorithm is as  follows 
({\it{for notational convenience the frequency arguments are not
displayed below}}):  
\begin{enumerate}
\item Make a guess for the impurity-averaged cluster self energy 
${\bar{\Sigma}}(\K)$, usually zero.
\item Calculate the coarse-grained Cluster Green functions
\begin{equation}
{\bar{G}}(\K) = \frac{N_c}{N} \sum_\kt 
\frac{1}{\omega +\mu-\epsilon_{\K+\kt} - {\bar{\Sigma}}(\K)}
\end{equation}
\item Calculate the cluster-excluded propagator 
${\bf\cal{G(\K)}}=1/(1/{\bar{G}}(\K)+{\bar{\Sigma}}({\K}))$.  The introduction 
of ${\cal{G(\K)}}$ is necessary to avoid overcounting diagrams on the cluster.
\item Fourier transform ${\cal{G}}$ to the 
real space matrix representation of the cluster problem,
(i.e., write $ {\bf\cal{G}}_{n,m} = \sum_{\K} {{\bf\cal{G}}(\K){\exp{i\K
\cdot(\r_n-\r_m)}} }$  ) whence 
the disorder potential may be represented as a diagonal matrix ${\bf{V}}$
(with elements  $V_n$ on the cluster sites labelled by $n$)
and form a new estimate of the disorder-averaged cluster Green function
matrix
\begin{equation}
{\bf{G}} = \left < 
\left({\bf\cal{G}}^{-1} - {\bf{V}}\right)^{-1} \right>
\end{equation}
where the average $\left< \right>$ indicates an average over
disorder configurations on the cluster \cite{adv_dca}.
\item Transform back to the cluster reciprocal space and form a new estimate 
of the self energy ${\bar{\Sigma}}(\K) = 1/{\cal{G}}(\K)-1/G(\K)$
\item Repeat, starting from 2, until ${\bar{\Sigma}}(\K)$ converges to the
 desired accuracy.
\end{enumerate}
The algorithm recovers the CPA for $N_c=1$ and becomes exact when 
$N_c\to\infty$.

\section{Characteristics of the DCA}
\label{characteristics}

In Ref.~\cite{agonis} A.~Gonis discusses the CPA, and various methods
to incorporate non-local corrections.  He lists the most important 
characteristics of a successful cluster theory.  A successful theory must 
be able to account for fluctuations in the local environment in a 
self-consistent way, become exact in the limit of large cluster sizes, 
and recover the CPA when the cluster size becomes one.  It must 
be straight-forward
to implement numerically and preserve the full point-group symmetry of the 
lattice.  Finally, and most significantly, it should be fully causal so 
that the single-particle Green function and self energy are analytic in the 
upper half plane.  In the next two sections, we demonstrate that the
DCA satisfies each of these requirements.

\paragraph*{The limits $N_c\to 1$ and $N_c\to \infty$}:  
As mentioned above, the DCA recovers the CPA for $N_c=1$.  
When $N_c = 1$, $\K = 0$, and $\kt =\k$. Then the DCA algorithm reduces to
the self consistent {\em{scalar equations}} (in contrast to the DCA which
involves matrix equations): 
\begin{equation}
{\bar{G}} = \frac{1}{N} \sum_\k 
\frac{1}{\omega +\mu-\epsilon_{\k} - {\bar{\Sigma}} }, 
\end{equation}
\begin{equation}
{\bf\cal{G}}^{-1} =  1/{\bar{G}} + {\bar{\Sigma}} ,
\end{equation}
\begin{equation}
{\bar{G}} = \left< \left( {\bf\cal{G}}^{-1} - V \right)^{-1} \right>,
\end{equation}
which together correspond exactly to the prescription for CPA
\cite{agonis}, and are easily solved, for example by iteration.

As $N_c\to \infty$ the DCA becomes exact, including correlations over all
length scales.   For, the DCA Laue function requires complete momentum
conservation in  this limit, i.e., $[\K, \bar{G}(\K) ] \equiv [\k, G(\k)
]$, whence $ {\cal{G}}= 1/(\omega+\mu-\epsilon_\k) $ which is
the bare lattice Green function so that Eq.~12 amounts to solving the problem 
by exact diagonalization.

\paragraph*{Causality}:
It is easy to show that this algorithm is fully causal\cite{DCA_hettler2}.   
\begin{figure}[htb]
\epsfxsize=2.6in
\epsffile{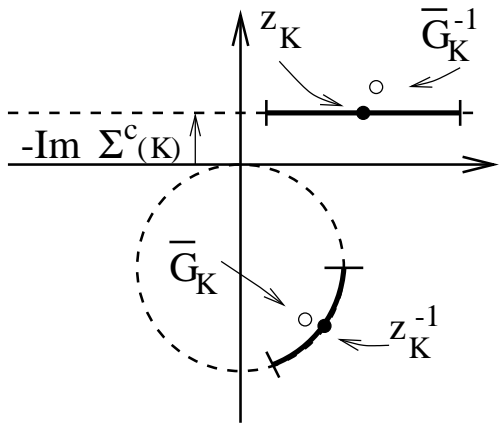}

\caption{Illustration of the essential steps of the proof that the DCA is
causal (see text).}
\label{causal}
\end{figure}
For our purposes, this is equivalent to requiring that all the retarded 
propagators and self energies are Herglotz, or analytic in the upper half 
plane.  Since the diagrams which describe the DCA cluster problem are 
isomorphic to those of a real finite-size periodic cluster, the 
corresponding impurity averaged cluster self energy shares the causal 
properties of this system.  Furthermore, the coarse graining step (2) cannot 
violate causality, since the sum of analytic functions is analytic.  The 
only ``suspect'' step is the cluster-exclusion step (3), thus we must show that 
the imaginary part of ${\cal{G}}$ is negative semidefinite.  The essential 
steps of the argument are sketched in Fig.~\ref{causal}.  The imaginary 
part of 
${\cal{G}}(\K,\omega)=(\bar{G}(\K,\omega)^{-1} + 
{\bar{\Sigma}}(\K,\omega))^{-1}$ is  negative provided that 
${\rm Im} (\bar{G}(\K,\omega)^{-1}) \geq -{\rm Im}{\bar{\Sigma}}(\K,\omega)$. 
$\bar{G}(\K,\omega)$ can be written as
$\bar{G}(\K,\omega)=(N_c/N) \sum_{\kt} (z_{\K+\kt})^{-1}(\omega)$, where
the $z_{\K+\kt}(\omega)$ 
are complex numbers with a positive semidefinite imaginary part 
$-\mbox{Im}{\bar{\Sigma}}(\K,\omega)$.
For any $\K$ and $\omega$, the set of points 
$z_{\K+\kt}(\omega)$ are on a segment of  the 
dashed {\it horizontal} line  in the upper half plane 
due to the fact that the imaginary part is {\it independent} of $\kt$.
The mapping $z\rightarrow 1/z$ maps this line segment onto a segment
of the dashed circle shown in the lower half plane. $\bar{G}(\K,\omega)$ 
is obtained by summing the points on the circle segment, yielding the 
empty dot that must lie {\it within} the dashed circle.  The inverse 
necessary to take $\bar{G}(\K,\omega)$ to $1/\bar{G}(\K,\omega)$ maps this
point onto the empty dot in the upper half plane which must lie {\it above}
the dashed line.  
Thus, the imaginary part of $\bar{G}(\K,\omega)^{-1}$ is greater than or 
equal to $-{\rm Im}{\bar{\Sigma}}(\K,\omega)$.  This argument may easily
be  extended for ${{\cal{G}}(z)}$ for any $z$  in the upper half plane.
Thus ${\cal{G}}$ is completely analytic in the upper half plane.

\paragraph*{Preservation of the Lattice Symmetry}:
Since the DCA is formulated in reciprocal space, it preserves the
translational symmetry of the system. However, care must be taken when
selecting the coarse-graining cells to preserve  the point group 
symmetries of the lattice.  For example, the calculations presented in 
Sec.~\ref{results} are done for a simple square  lattice. Both it and 
its reciprocal lattice have a $C_{4v}$ symmetry with  eight
point group operations.   We must choose a set of coarse-graining cells 
which preserve this point-group
symmetry.  This may be done by tiling the real 
lattice with squares, and using the $\K$ points that correspond to the 
reciprocal space of the tiling centers.  For large $V$, an important 
configuration of the half filled system is that with all of the sites 
with $V_i=-V$ occupied and those with $V_i=+V$ unoccupied.  To retain  this
configuration on the cluster, when $N_c>1$ we will choose $N_c$ even.   

\begin{figure}[htb]
\epsfxsize=3.25in
\epsffile{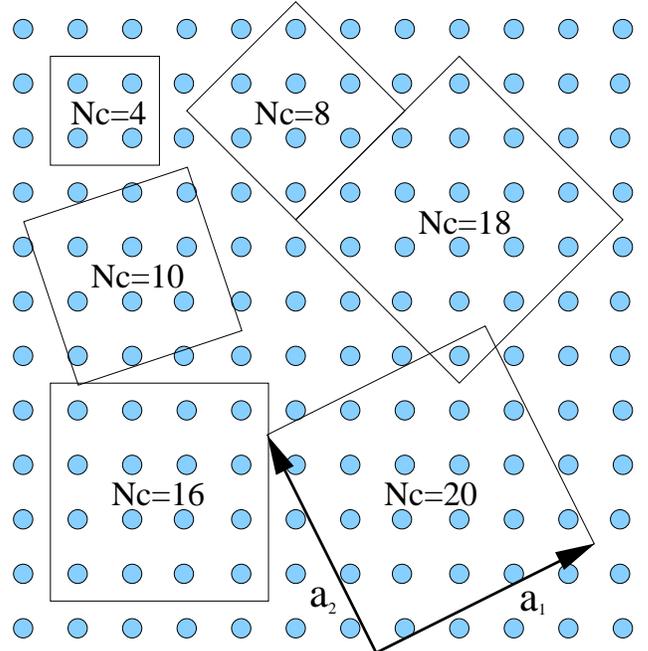}

\caption{Different tile sizes and orientations.  The tiling principal
translation vectors, $\a_1$ and $\a_2$, form two sides of each tiling 
square (illustrated for the $N_c=20$ tiling).  For square tile 
geometries, $a_{2x}=-a_{1y}$ and $a_{2y}=a_{1x}$.
}
\label{tilings}
\end{figure}
\begin{figure}[htb]
\epsfxsize=3.25in
\epsffile{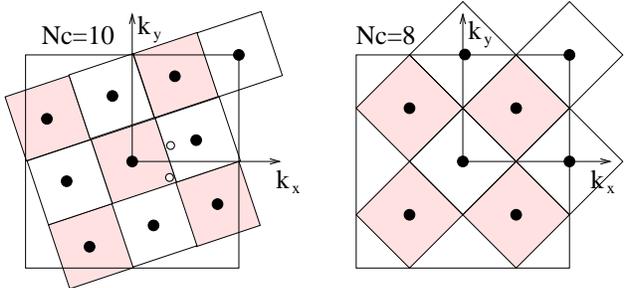}

\caption{The coarse graining cells for $N_c=8$ and $10$ each centered on
a coarse-grained momenta $\K$ represented as black filled dots.  For $N_c=8$
equivalent momenta $\k$ are always mapped to equivalent coarse-grained
momenta $\K$.  However, this is not true for $N_c=10$ where, for example,
the two equivalent momenta shown by open dots are mapped to inequivalent
coarse-grained momenta.}
\label{Nc8_10}
\end{figure}

Square tilings with an even number of sites include $N_c=4,8,10,16,18,20,
26,32,34,36,\cdots$.  The first few are illustrated in Fig.~\ref{tilings}.  
The relation between the principal lattice vectors of the lattice centers, 
$\a_1$ and $\a_2$, and the reciprocal lattice takes the usual form 
$\g_i = 2\pi\a_i/\left|\a_1\times\a_2\right|$, with $\K_{nm}= n\g_1+m\g_2$ 
for integer $n$ and $m$.  For tilings with either $a_{1x}=a_{1y}$ 
(corresponding to $N_c= 8,18,32\cdots$) or one of $a_{1x}$ or $a_{1y}$ zero 
(corresponding to $N_c=1,4,16,36\cdots$), the principal reciprocal lattice 
vectors of the coarse-grained system either point along the same directions 
as the principal reciprocal lattice vectors of the real system or are rotated 
from them by $\pi/4$.   As a result equivalent momenta $\k$ are always mapped 
to equivalent coarse-grained momenta $\K$. An example for $N_c=8$ is shown 
in Fig.~\ref{Nc8_10}.  However, for $N_c=10,20,26,34\cdots$, the principal 
reciprocal lattice vectors of the coarse-grained system do not point along a 
high symmetry direction of the real lattice.  Since all points within a 
coarse-graining cell are mapped to its center $\K$, this means that these 
coarse graining choices violate the point group symmetry of the real system.  
This is illustrate for $N_c=10$ in Fig.~\ref{Nc8_10}, where the two open dots, 
resting at equivalent points in the real lattice, fall in inequivalent coarse 
graining cells and so are mapped to inequivalent $\K$ points.  Thus the 
tilings corresponding to $N_c=10,20,26,34\cdots$ violate the point-group 
symmetry of the real lattice system and should be avoided.

\paragraph*{An efficient numerical algorithm for disorder averaging}:
The implementation of the DCA clearly requires an efficient algorithm for 
disorder averaging on a cluster of size $N_c$.( Needless to say, this
particular aspect is common to all approaches where disorder averaging is
involved.) Even for a system with binary  diagonal disorder, where each
site can acquire only one of two values for  the potential ($\pm V$) the
total number of disorder configurations is  $2^{N_c}$, which grows 
exponentially with $N_c$.  For  a generic quenched disorder, the 
probability of the various disorder configurations are determined by the
specified $P(V)$; whereas, for annealed disorder, the probability of a 
configuration depends upon an effective Boltzmann factor determined by 
the electronic partition function for that particular disorder 
configuration \cite{DCA_hettler2}.

In this  section, we propose an approach to carrying out the disorder 
averaging by statistically sampling disorder configurations using a 
Markov process. For systems with quenched disorder, one could also sample 
random configurations of the disorder potentials and calculate the  
corresponding Green function using matrix inversion. A significant  
advantage of the Markov technique is that it may be easily modified to 
treat  either quenched or annealed disorder\cite{DCA_hettler2}.

	In a Markov process, each disorder configuration depends upon the
previous configurations.  To evolve from one configuration to another, we
will  propose local changes in the disorder potentials.  These changes are
accepted with some probability determined by either the effective Boltzmann 
factor \cite{DCA_hettler2}, for annealed disorder, or the probability 
distribution $P(V_i)$ , for quenched disorder.  If we accept such a change 
in one of the disorder potentials, say on site $l$, then the new Green 
function matrix $G'_{n,m}$ depends on the previous Green function matrix 
$G_{n,m}$ through the matrix relationship
({\it{where, once again, for notational convenience the frequency arguments
are not displayed}})
\begin{equation}
{\bf{G'}}^{-1}-{\bf{G}}^{-1} = {\bf{V}}-{\bf{V}}'
\label{G1}
\end{equation}
Since we change on the potential on the site $l$ and the matrices
${\bf{V}}$ and ${\bf{V}}'$ are diagonal, their 
difference $\delta{\bf{V}}={\bf{V}}-{\bf{V}}'$ 
is a diagonal matrix with only the $l$'th diagonal element finite.
Then 
\begin{equation}
G'_{n,m} = G_{n,m} + G_{n,l} \delta V_l G'_{l,m}
\label{G2}
\end{equation}
If we set $l=n$ in Eq.~\ref{G2}, we get
\begin{equation}
G'_{l,m} = \frac{G_{l,m}}{1 - G_{l,l} \delta V_l}
\label{G3}
\end{equation}
If we substitute this result back into Eq.~\ref{G2}, we get
\begin{equation}
G'_{n,m} = G_{n,m} + 
G_{n,l} \frac{\delta V_l}{1 - G_{l,l} \delta V_l} G_{l,m}
\end{equation}
an equation which requires roughly $N_c^2$ operations to evaluate
\cite{hirsch_fye} for each frequency.
	
Because the technique involves importance sampling, it is likely
to miss rare configurations of disorder, and any special physics that
arises from such configurations, such as Lifshitz tails in the DOS.
However, when these configurations are well known,  we can easily adapt 
this method to include them.  This may be done by excluding them from 
the sampling, and then including the corresponding configurations in
the sample with the appropriate reweighting.

\section{Results}
\label{results}

\subsection{Single-Particle Properties}

	To illustrate the algorithm discussed above, we present
calculations on a two-dimensional square lattice system with 
$t=0.25$ and binary random disorder so that $V_i=\pm V$ 
with equal probability.  No tricks are used to force the algorithm 
to remain causal such as renormalizing the spectrum or cutting off any
negative tails of the density of states.

\begin{figure}[htb]
\epsfxsize=3.3in
\epsffile{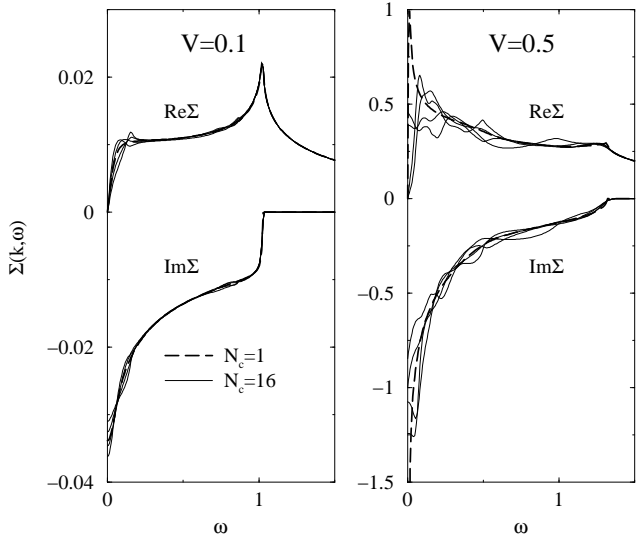}

\caption{The self energy $N_c=1$ and for $N_c=16$ for different values 
of $\K$ for $V=0.1$ (left) and $V=0.5$ (right).  The thin lines show
${\bar{\Sigma}}(\K,\omega)$ for $\K=(0,0)$ $\K=(0,\pi)$, $\K=(\pi/2,\pi/2)$ and
$\K=(\pi/2,0)$, plotted versus $\omega$ for $N_c=16$.  For $V=0.1$
the self energy at the different values of $\K$ is essentially the
same as the $N_c=1$ result.  For $V=0.5$ they differ significantly,
indicating that the momentum dependence of the self energy increases
with $V$.
}
\label{Sigmas}
\end{figure}

	The self energy is plotted in Fig.~\ref{Sigmas} for $N_c=1$ 
and for $N_c=16$ for different values of $\K$ for $V=0.1$ (left) and 
$V=0.5$ (right).  For $V=0.1$ the self energy for $N_c=16$ has very
little momentum dependence, thus the different curves fall atop
of one another.  They also are hence very close to the self energy 
for $N_c=1$ (the CPA result) indicating that it is a very good
approximation for the self energy when $V$ is small.  However, for 
larger $V=0.5$ the self energy curves at different values of $K$ for 
$N_c=16$ differ considerably from each other and from the CPA self energy
obtained with $N_c=1$.  Thus, as $V$ increases non-local corrections
clearly become important and are expressed in the momentum-dependence of
the self energy.

\begin{figure}[htb]
\epsfxsize=3.25in
\epsffile{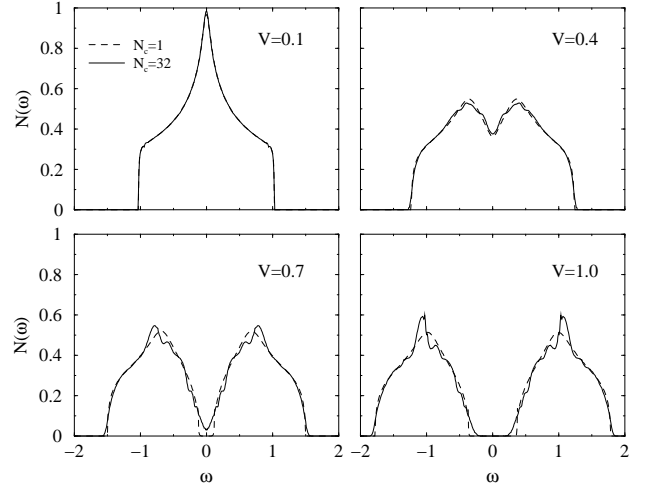}

\caption{The density of states for $N_c=1$ and for $N_c=32$ for four 
values of disorder potential $V$.
}
\label{DOS_compare}
\end{figure}

	As shown in Fig.~\ref{DOS_compare} (top left) the single particle
density of states is essentially independent of $N_c$ for $V=0.1$;
however, for $V=1.0$, the density of states depends strongly upon 
$N_c$.  In particular the gap around $\omega = 0$, which is sharp 
for $N_c=1$, is partially filled in.  The top and bottom of the band 
also acquire tails as $N_c$ increases.  When $N_c>1$, the density of 
states acquires several additional structures which correspond to
important local configurations of the disorder.  The additional features
and the band tails are absent in the CPA and believed to be due to
local order in the environment of each site\cite{agonis}.

\subsection{(Absence of) Localization}

Despite its advantages over the CPA as discussed above, one feature that
the DCA shares with the CPA and similar self-consistent cluster methods 
is its limited ability to take into account localization effects
\cite{review1,McKane_Stone,review2}. 
To show this, we measure the probability  that an electron remains at 
site $l$ for all time\cite{review1,McKane_Stone}: 

\begin{eqnarray}
P(\infty) &=& \lim_{t\to\infty} \left< \left|G(l,l,t)\right|^2\right>
\nonumber \\
&=&\lim_{\eta\to 0} \frac{\eta}{\pi}\int_{-\infty}^\infty d\epsilon
\left< \left|G_{l,l} (\epsilon+i\eta)\right|^2\right>\,.
\label{localize_test1}
\end{eqnarray}

\begin{figure}[htb]
\epsfxsize=3.25in
\epsffile{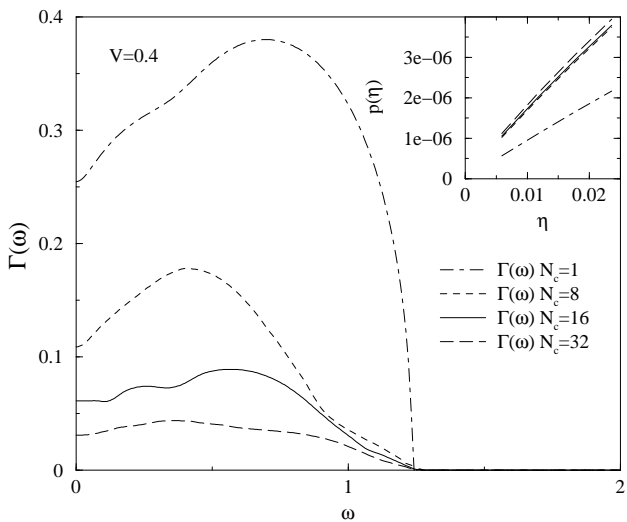}

\caption{The local hybridization rate when $V=0.4$ for several values of 
$N_c$.  The probability that the electron remains localized $p(\eta\to 0)$, 
when $\epsilon=0$ (half filling) is shown in the inset.  As $\eta\to 0$, 
$p(\eta)$ extrapolates to zero indicating the lack of localization.
}
\label{Gamma}
\end{figure}

As shown in\cite{review1,McKane_Stone}, $P(\infty)$ is expected to be 
nonzero as long as there are a thermodynamically significant fraction 
of localized states in the spectrum of eigenstates of the disordered 
system.  In 1 and 2 dimensions this is expected to happen for 
arbitrarily small but thermodynamically significant disorder.  Since 
the cluster is formed by coarse-graining the real-lattice problem in 
reciprocal space, local quantities on the cluster and the real lattice 
correspond one-to-one.  Thus, to test for localization, we need only 
apply  the formula \ref{localize_test1} for each site on the cluster.  
Making this substitution and introducing the local coarse-grained (but 
not disorder averaged) spectral function, 
${\bar{A}}(l,\omega)= -\frac{1}{\pi}{\rm{Im}}{\bar{G}}_{l,l}(\omega)$,
Eq.~\ref{localize_test1} becomes  \begin{eqnarray} P(\infty) &=&
\lim_{\eta\to 0} p(\eta)  \nonumber \\ &=&\lim_{\eta\to 0}
\frac{-2i\eta}{N_c} \sum_l \int_{-\infty}^\infty  d\omega d\omega'
\left< \frac{{\bar{A}}(l,\omega){\bar{A}}(l,\omega')}
{\omega-\omega'-2i\eta} \right>\,.
\label{localize_test2}
\end{eqnarray}
$p(\eta)$ is plotted versus $\eta$ in the inset to Fig.~\ref{Gamma} for the 
half-filled model when $V=0.4$.  The $p(\eta)$ extrapolates to zero, 
indicating the lack of localization.

This result can be understood from either a diagrammatic perspective or by
carefully assessing the cluster problem.  As is well known \cite{review2},
the crossing diagrams, especially  those which involve many crossings,
describe the coherent backscattering of electrons which are responsible
for localization. Hence the CPA, which includes only non-crossing
diagrams, can not describe localization.  Within the DCA, however, for 
$N_c>1$ {\em{some crossing graphs are restored }}.  Within each diagram,
each $X$  represents scattering from distinct site.  Since  there are only
$N_c$ sites  on the cluster, the maximally crossed DCA graphs can have at
most $N_c$ crossings. Since all states are expected to be localized in two
dimensional disordered system, apparently an infinite number of crossings
are needed to describe localization diagrammatically.  From the
perspective of the cluster, this result is not surprising since each site
on the cluster is coupled to a non-interacting translationally invariant
host into which electrons can escape.  Thus, if the density of states is
finite at some energy, then the corresponding states can not be localized
unless the hybridization rate at that energy between the cluster and the
host vanishes. 
As described in Ref.~\cite{DCA_maier1}, the hybridization rate between the 
cluster and its host is given by
\begin{equation}
\Gamma(\K,\omega) = 
{\rm{Im}} \left( \frac{1}{{\bar{G}}(\K,\omega)} + {\bar{\Sigma}}(\K,\omega)\right)
\end{equation}
The net hybridization rate to a site on the cluster (the $\K$-integrated
$\Gamma(\K,\omega)$) is plotted in Fig.~\ref{Gamma} when $V=0.4$ for several 
values of $N_c$.  It remains finite over the entire region where the 
corresponding density of states, shown in Fig.~\ref{DOS_compare}, is finite.  
This is consistent with the lack of localization demonstrated in the inset.

\begin{figure}[htb]
\epsfxsize=3.25in \epsffile{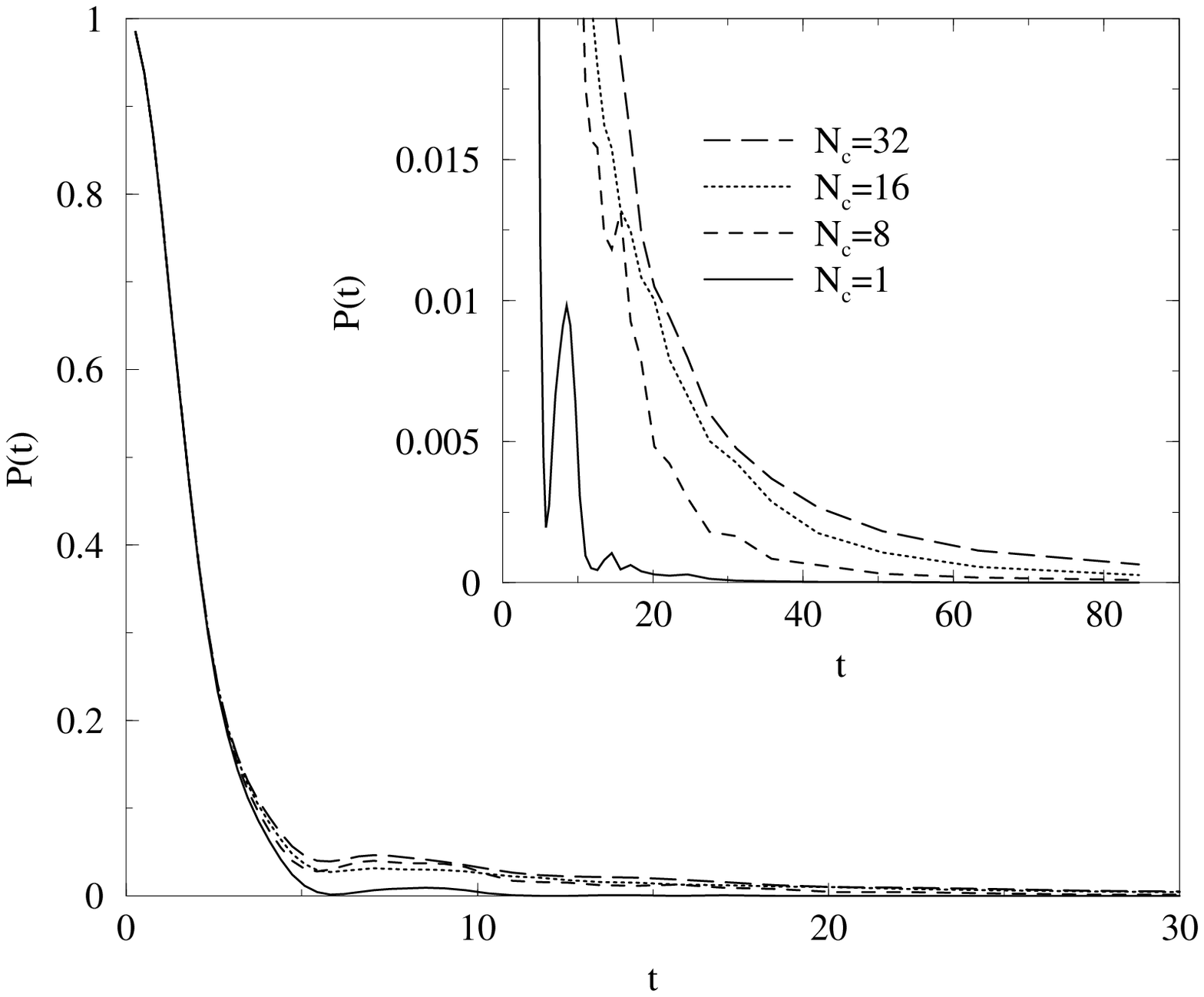}

\caption{The probability that an electron on a site $l$ remains after 
a time $t$, $P(t) =  \left< \left|G(l,l,t)\right|^2\right>$ for several
values of $N_c$ when $V=0.4$.
}
\label{P(t)}
\end{figure}

We note that the hybridization falls as $N_c$ increases (for large $N_c$, 
$\Gamma(\K,\omega)\sim {\cal{O}}(1/N_c)$ \cite{DCA_maier1}) especially
at the band edges, although, given that it is defined entirely in terms of 
disorder-averaged propagators, it is still unlikely to be sensitive to 
localization effects.   However, the number of diagrammatic crossings in 
two particle properties (such as the conductivity) which are strongly
affected by localization effects does increase with $N_c$ even within 
the DCA. 
Thus, it is likely that disordered DCA can describe the precursor effects
of localization.  Some evidence for this can be seen in
the (finite time) probability that an electron on a site $l$ remains after
a time $t$, $P(t) =  \left< \left|G(l,l,t)\right|^2\right>$.  As shown in 
Fig.~\ref{P(t)}, for $N_c=1$, this probability falls quickly with time.  
The long time behavior is shown in the inset.  As  $N_c$ increases, the 
electron remains localized for longer times. Hence one can hope that a
careful finite size scaling study of two particle properties within the
disordered DCA can even capture some aspects of the localization
transition.

\section{Conclusion}

	We have developed a modification of the Dynamical Cluster 
Approximation to treat disordered systems.  This formalism satisfies all 
of the characteristics of a successful cluster approximation.  It is 
causal, preserves the point-group and translational symmetry of the 
original lattice, recovers the CPA when the cluster size goes to one, 
and becomes exact as $N_c\to\infty$.  Like the CPA the problem is 
disorder averaged and has a simple diagrammatic formulation. It is easy 
to implement numerically and restores sharp features and band tailing 
in the DOS which reflect correlations in the local environment of each 
site.   Although the DCA does not capture the localization transition, 
it does describe the precursor effects of localization. It 
systematically restore the crossing graphs known to be responsible for 
localization,  and might be able to access the localization transition 
itself via an appropriate finite size scaling analysis of two particle 
properties which remains to be developed. 

	The DCA formalism we have discussed here can also be extended to
problems  with {\em{disorder and interactions}} simply by incorporating
interaction  diagrams in the self energy.  This is also discussed in the
Appendix below. The DCA should be able to provide a good description of 
localization effects at finite temperatures in such contexts.  For, in 
such cases the scattering processes are partially inelastic, so that the 
coherent back scattering disappears after a characteristic inelastic 
scattering time\cite{berg1984}.  In this time only a finite-number of 
back-scattering processes can occur so only a finite number of 
diagrammatic crossings  are needed to describe the finite-temperature 
physics, and these are captured in the DCA.

\paragraph*{Acknowledgements} 
This work was initiated in conversations with B.L.~Gyorffy.  It is a 
pleasure to acknowledge useful discussions with 
F.P.~Esposito,
A.~Gonis,
M.~Hettler, 
D.~E.~Logan,
and
M.~Ma
. This work was supported in part by NSF grants DMR-9704021, 
DMR-9357199, and PHY94-07194 and by PRF grant ACF-PRF\#33611-AC6.
This research was supported in part by NSF cooperative agreement 
ACI-9619020 through computing resources provided by the National 
Partnership for Advanced Computational Infrastructure at the San 
Diego Supercomputer Center.

\appendix

\section{Disorder-DCA from the Replica Method}

	An alternate way of justifying the DCA in the context of
disordered systems is to use the replica (or other such) trick for disorder
averaging \cite{replica}. For, this maps the disorder averaged problem
into what looks like an interacting problem, whence the DCA formalism
developed by us earlier \cite{DCA_hettler1,DCA_hettler2} can simply be
transcribed for this case, to arrive at the appropriate self consistent
cluster problem. For the effective cluster problem, the replica trick can
be "un-done", and we recover the algorithm presented earlier in this
paper. The same procedure also works for problems involving {\em{both
disorder and interactions}}. We detail this below.

	As is well known, for problems involving {\em{quenched}} disorder,
as for example  corresponding to the Hamiltonian:
\begin{eqnarray}
H & = & H_0 + H_{dis}\\
H_0 & = & \sum_{\k,\sigma} \xi_{\k} C^{\dagger}_{\k,\sigma}
C_{\k,\sigma}\\  
H_{dis} & = & \sum_{i \sigma} V_i n_{i,\sigma}
\end{eqnarray}
where $ \xi_{\k} = \epsilon_{\k} - \mu $, and $ V_i $ is the random
potential distributed according to a given probability distribution $P(V)$,
complications arise because the disorder averaging has to be done  on the
{\em{free energy}} 
$$ F = -k_B T \ln Z $$
and the {\em{green functions}}
$$ G_{i,j} (\tau) = -Tr [ {\bf{T}}_\tau C_{i,\sigma}(\tau)
C^{\dagger}_{j,\sigma} exp( - \beta H ) ] / Z .$$
Here $ Z = Tr  [ exp( - \beta H ) ] $ is the partition function, and 
${\bf{T}}_\tau$ represents the imaginary-time ordering operator.

	In the replica trick \cite{replica}, one writes 
$$ \ln Z = \lim_{m_r \to 0} \frac { Z^{m_r} - 1} { m_r}
\mbox{ and } 1/Z = \lim_{m_r \to 0} Z^{m_r - 1} $$
and {\em{assumes}} that the order of taking the limit ${m_r \to 0}$ and
disorder averaging can be interchanged. Then for any positive integer
$m_r$, the resulting disorder averaged quantities such as
 $ \left< Z^{m_r} \right> $ , $ \left< G_{\k}(\tau) \right> $, etc., can be
represented in terms of an {\em{interacting}} problem involving $m_r$
replicas of the original electronic degrees of freedom, which we index
with the subscript $\alpha = 1,...,m_r $. 

	For example, using the standard Fermionic (Grassmann
variable) functional integrals \cite{grass} to represent the traces above,
we can write
\begin{eqnarray}
\left< Z^{m_r} \right> & = & \int D\, c^* D\, c \; \; 
\exp{ [-\beta\Psi ] } \\ 
\left< G_{\k}(\tau) \right> & = & - \int D\, c^* D\, c \; \;
c_{\k,\sigma,1}(\tau) \; c^*_{\k,\sigma,1}(0) \; 
\exp{ [ -\beta \Psi ] } 
\end{eqnarray}
Here $\Psi$ is an effective free energy functional which arises from the
disorder averaging, and can be written as 
\begin{equation} 
\beta \Psi = \sum_{\k,\sigma,\alpha} \int_0^\beta d\tau \; c^*_{\k,\sigma,
\alpha}(\tau)  (\partial_\tau + \xi_{\k} )  c_{\k,\sigma ,\alpha}(\tau)
+ \sum_{i}^{N} W ( \tilde{n}_i )\\
\label{psi-latt}
\end{equation}
where,
$$ \tilde{n}_i \equiv \sum_{\alpha,\sigma} \int_0^\beta n_{i,\sigma,\alpha}
(\tau) \; d\tau $$
and 
$$ \exp{ [-W ( \tilde{n}_i)] } = \left< \exp{( - V_i \tilde{n}_i) }
\right>  = \int { dV_i P(V_i) exp{ (- V_i \tilde{n}_i) } } $$
In terms of the cumulants $ {\left< V^l \right>}_c $ of the disorder
distribution $ P(V) $, one can write
$$ W ( \tilde{n}_i) = \sum_{l=2} ^{\infty} \frac {1} {l!}
{\left< V^l \right>}_c (\tilde{n}_i)^l $$
So, clearly, W introduces (local in space but non-local in time)
interactions between electrons belonging to arbitrary replicas.

	If one re-expands $\exp{ [-W ( \tilde{n}_i)] } $ in powers of the
cumulants $ {\left< V^l \right>}_c $ one can perform the Fermionic traces
using the standard techniques of diagrammatic perturbation theory. Then,
order-by-order in perturbation theory, the dependence on $m_r$ is explicit
and analytic, and the $\lim_{m_r \to 0}$ can be evaluated precisely. The
resulting terms are in exact, one-to-one correspondence with the terms
obtainable by writing out the diagrams from a direct perturbation
expansion in powers of $V_i$ and then disorder averaging as discussed in 
Section II. The ${m_r \to 0}$ limit eliminates the diagrams (in the
interacting problem ) containing internal loops with free sums over the
replica indices (as required, since such diagrams never appear in the
direct disorder averaged perturbation theory formalism of Section II ).

	For the "replicated interacting problem" obtained above, one can
transcribe exactly the DCA formalism discussed in 
refs.~\cite{DCA_hettler1,DCA_hettler2}. If one assumes that the
self-consistent host propagators do not break replica symmetry, then the
effective cluster problem corresponds to a Fermionic functional
integral involving an effective, self consistent cluster free energy
functional given by 
\begin{eqnarray} 
\beta {\Psi}_c & = & \sum_{\K,\sigma,\alpha} \int_0^\beta d\tau
\int_0^\beta d{\tau}' \; c^*_{\K,\sigma,\alpha}(\tau) 
{\bf{\cal{G}}}^{-1}(\K, \tau - {\tau}')  c_{\K,\sigma,\alpha}({\tau}')
\nonumber \\      
& & + \sum_{i}^{N_c} W ( \tilde{n}_i )
\label{psi-clust}
\end{eqnarray}
But, as is easy to see using the same procedure as outlined earlier in
this appendix, such an effective free energy functional is exactly what one
would obtain if one were to disorder average (using the replica trick) a
cluster problem with $N_c $ sites which are dual to the cluster momenta
$\K$, a bare ( retarded ) cluster propagator $\bf{\cal{G}}^{-1}(\K, \tau -
{\tau}')$, and a random potential $V_i$ distributed according to $P(V_i)$
at every site $i$ of the cluster. Hence we have an alternate justification
for the disorder-DCA algorithm set down in Section III. The above route 
also enables one to quickly extend our discussions in ref.\cite{DCA_hettler2}
regarding the 2-particle propagators, Ward identities, etc., to the
disorder-DCA context.  	 
Most significantly, the DCA
estimate of the lattice self energy is minimized by the choice
$\Sigma_\alpha(\k,\omega) =  {\bar{\Sigma}}(\M(\k),\omega)$.
	 
	We note that the arguments presented in the main text and in this 
appendix are also easily extended to problems involving {\em{interactions 
and disorder}}. For example, for the case of the Hubbard model with diagonal 
disorder, one would add to the starting Hamiltonian the interaction term  
$U \sum_{i}^{N} n_{i,\uparrow} n_{i,\downarrow} $. Going through exactly
the same procedures as outlined above, it is not hard to see that the only
change is that the effective free energy functionals for the lattice and
the cluster pick up the additional terms 
$U \sum_{\alpha} \sum_{i^N} n_{i,\uparrow,\alpha}
n_{i,\downarrow,\alpha}$ and 
$U \sum_{\alpha} \sum_{i^{N_c}}
n_{i,\uparrow,\alpha} n_{i,\downarrow,\alpha}$. The resulting cluster
problem now has both interactions and disorder on the cluster of $N_c $
sites which are dual to the cluster momenta $\K$: a bare ( retarded )
cluster propagator $\bf{\cal{G}}^{-1}(\K, \tau - {\tau}')$, a random
potential $V_i$ distributed according to $P(V_i)$, and the Hubbard
interaction U at every site $i$ of the cluster. One can resort to  any
technique of one's choice to solve this problem for the  disorder-
averaged cluster Green functions $\bar{G}(\K,\omega)$ and  cluster
self-energies ${\bar{\Sigma}}(\K,\omega)$ and go through with the  rest of
the DCA iteration.  


\begin{thebibliography}{99}
\bibitem{fntca} In particular, what are arguably the best of these
extensions, namely the TCA approach \cite{mlsrv78} and its generalizations
\cite{kldg80} are designed to be causal, but are not systematic, in that
the terms which are retained by them may be just as important as those
dropped. Furthermore they are so hard to implement numerically that this
has never been done for clusters larger than two sites.
\end{thebibliography}
\end{document}